\documentclass[12pt]{article}
\pdfoutput=1

\usepackage[utf8]{inputenc}
\usepackage[tbtags]{amsmath} 				
\usepackage{amssymb, mathrsfs}			
\usepackage{physics}
\usepackage{graphicx}			
\usepackage{tensor}				
\usepackage[makeroom]{cancel}	
\usepackage{bm}								
\usepackage[
      colorlinks=true,
      linkcolor=blue,
      urlcolor=blue,
      filecolor=black,
      citecolor=red,
      pdfstartview=FitV,
      pdftitle={},
      pdfauthor={Michael Gutperle, Kevin Chen, Charlie Hultgreen-Mena},
      bookmarksopen=true
      ]{hyperref}
\usepackage{calc}
\usepackage{sectsty}
\usepackage{dsfont}
\usepackage{cite}

\newlength\myheight
\newlength\mydepth
\settototalheight\myheight{Xygp}
\DeclareMathOperator{\diag}{diag}

\DeclareMathOperator{\SO}{SO}

\DeclareMathOperator{\OSp}{OSp}

\newcommand{\AdS}{\text{AdS}}
\newcommand{\cA}{\mathcal A}
\newcommand{\cB}{\mathcal B}

\newcommand{\cK}{\mathcal K}

\newcommand{\cP}{\mathcal P}
\newcommand{\cN}{\mathcal N}
\newcommand{\cM}{\mathcal M}

\newcommand{\cV}{\mathcal V}
\newcommand{\cQ}{\mathcal Q}
\newcommand{\cL}{\mathcal L}

\sectionfont{\large}

\renewcommand{\thefootnote}{\fnsymbol{footnote}}
\renewcommand{\thanks}[1]{\footnote{#1}}
\newcommand{\starttext}{
\setcounter{footnote}{0}
\renewcommand{\thefootnote}{\arabic{footnote}}}
\renewcommand{\epsilon}{\varepsilon}	

\numberwithin{equation}{section} 		

\numberwithin{equation}{section}

\long\def\symbolfootnote[#1]#2{\begingroup%
\def\thefootnote{\fnsymbol{footnote}}\footnote[#1]{#2}\endgroup}

\begin{document}
\setlength{\baselineskip}{16pt}

\starttext
\setcounter{footnote}{0}

\begin{flushright}
\today
\end{flushright}

\bigskip

\begin{center}

{\Large \bf  Janus  and RG-flow interfaces in three-dimensional  gauged supergravity}

\vskip 0.4in

{\large  Kevin Chen, Michael Gutperle, and  Charlie Hultgreen-Mena }

\vskip 0.2in

{\sl Mani L.~Bhaumik Institute for Theoretical Physics}\\
{\sl Department of Physics and Astronomy }\\
{\sl University of California, Los Angeles, CA 90095, USA}

\end{center}
 
\bigskip
 
\begin{abstract}
\setlength{\baselineskip}{16pt}

In this paper, we construct Janus-type solutions of three-dimensional gauged supergravity with sixteen supersymmetries. We find solutions which correspond to interfaces between the same CFT on both sides,  as well as RG-flow interfaces between  CFTs with different numbers of supersymmetries and central charges.  The solutions are obtained by solving the flow equations derived from  the supersymmetry variations, and they preserve some fraction of the supersymmetries of the $\AdS_3$ vacua.
\end{abstract}

\setcounter{equation}{0}
\setcounter{footnote}{0}

\newpage

\section{Introduction}

Janus solutions provide holographic constructions of interface conformal field theories (CFTs).  In many known examples, the solutions are constructed by considering  an $\AdS_d$ slicing of a $(d+1)$-dimensional space where the scalar fields depend non-trivially on the slicing coordinate.  One of the most well-known examples is the Janus solution of \cite{Bak:2003jk},  which is a deformation of the $\AdS_5\times S^5$ vacuum of type IIB and is given by an $\AdS_4$ slicing where the dilation depends non-trivially on the slicing coordinate. The solution is dual to an interface of $N=4$ super Yang-Mills (SYM) theory where the coupling $g_{\rm YM}$ jumps across a co-dimension one interface \cite{Clark:2004sb}. This solution breaks all the supersymmetries, but a general solution given by an $\AdS_4 \times S^2 \times S^2$ space warped over a Riemann surface can preserve half the supersymmetries of the $\AdS_5\times S^5$ vacuum \cite{DHoker:2007zhm} and is dual to supersymmetric interface theories in $N=4$ SYM \cite{DHoker:2006qeo, Gaiotto:2008ak, Gaiotto:2008sd}. For other examples of Janus solutions in type II and M-theory, see e.g.~\cite{DHoker:2006vfr,DHoker:2009lky,DHoker:2007hhe}.

Instead of constructing solutions in ten or eleven dimensions, it is often useful to use lower-dimensional gauged supergravities since the ansatz and resulting equations are simpler. Often the resulting solutions can be uplifted to higher dimensions, but even if the uplift is not known the gauged supergravity solutions are useful for studying universal and qualitative features of interface solutions. For an incomplete list of such solutions in various dimensions, see e.g.~\cite{Pilch:2015dwa,Gutperle:2017nwo,Suh:2011xc,Bobev:2013yra,Clark:2005te,Suh:2018nmp,Karndumri:2016tpf,Karndumri:2020bkc}.

A related construction is given  by holographic RG-flows, which consider a Poincar\'e slicing instead of an $\AdS$ slicing. If the solutions asymptotically approach two AdS vacua with different cosmological constants, we can interpret this solution as an RG-flow from a  CFT in the UV to a CFT in the IR which is triggered by turning on a relevant deformation in the UV \cite{DeWolfe:1999cp,Freedman:1999gp,Skenderis:1999mm,Girardello:1999bd,Bianchi:2001de}.
On the other hand,  for most examples of AdS-sliced  holographic interface solutions  the CFTs on both sides of the interface are the same and differ only by a marginal  deformation (such as different values of $g_{\rm YM}$ in the example discussed above) or a position-dependent profile of the expectation value \cite{DHoker:2009lky} or a source \cite{Arav:2020obl} of a relevant operator.  

In \cite{Brunner:2007ur,Gaiotto:2012np},  the idea of a RG-interface in two-dimensional CFTs was proposed, where the two sides of the interface are CFTs related by a RG-flow. The goal of the current paper is to construct holographic solutions which realize this idea.\footnote{See \cite{Gutperle:2012hy,Arav:2020asu,Korovin:2013gha,Bobev:2013yra} for work  in this direction.}  We consider three-dimensional $\cN=8$ gauged supergravity with $n=4$ vector multiplets,  first discussed in \cite{Nicolai:2001ac}.  This theory has an $\AdS_3$ vacuum with maximal $\cN=(4,4)$ supersymmetry as well as two families of $\AdS_3$ vacua with $\cN=(1,1)$ supersymmetry \cite{Berg:2001ty}.  The theory gauges a $\SO(4)\times \SO(4)$ symmetry of the $\SO(8,4)/\SO(8)\times \SO(4)$ coset. The gauging  depends on a continuous parameter $\alpha$ where the  superconformal  algebra of the $\cN=(4,4)$ vacuum is given by the ``large'' superconformal algebra $D^1(2,1;\alpha) \times D^1(2,1;\alpha)$, and the three-dimensional supergravity is believed to be a truncation  of M-theory on $\AdS_3\times S^3\times S^3 \times S^1$ \cite{Boonstra:1998yu,Elitzur:1998mm,deBoer:1999gea,Gukov:2004ym}. In this paper, we primarily focus on the case $\alpha=1$ for which the expressions for the flow equations are the simplest.

In \cite{Berg:2001ty}, the  Poincar\'e-sliced holographic RG-flow solutions  were constructed for the $\cN=8$ gauged supergravity, which interpolate between the $\cN=(4,4)$ and $\cN=(1,1)$  vacua. 
The goal of the present paper is to construct Janus solutions which realize the interface between the same CFT at different points in the moduli space as well as RG-flow interfaces between CFTs preserving different numbers of supersymmetries.

The structure of this paper is as follows. In section \ref{sec2},  we review the $\cN=8$ gauged supergravity theory we will be using in the paper and discuss the different supersymmetric $\AdS_3$ vacua.  In section \ref{sec3},  we derive the BPS equations for a Janus ansatz following from the vanishing of the gravitino and spin-$\frac{1}{2}$ supersymmetry variations.  In section \ref{sec4}, we solve the BPS equations for various truncations which make the analysis of the flow equations manageable. For a Janus interface between the $\cN=(4,4)$ vacuum we find an analytic solution, whereas for the  RG-flow interfaces  the flow equations  can only be solved numerically. We present the solutions and provide evidence for our interpretation of these solutions as RG-flow interfaces.  In section \ref{sec5}, we close with a discussion and open questions. In appendix \ref{appendix-gamma}, we present our  conventions for $\SO(8)$ gamma matrices.

\section{Three-dimensional $\cN=8$ gauged supergravity}\label{sec2}
In this section, we review the  $\cN=8$ gauged supergravity first constructed in  \cite{Nicolai:2001ac}.   The theory is characterized by the number $n$  of vector multiplets.	The bosonic field content consists of a graviton $g_{\mu\nu}$, Chern-Simons gauge fields $B^\cM_\mu$, and scalars fields living  in a $G/H = \SO(8, n)/\SO(8) \times \SO(n)$ coset, which has $8n$ degrees of freedom before gauging.
The scalar fields can be parametrized by a $G$-valued matrix $L(x)$ in the vector representation, which transforms under $H$ and the gauge group $G_0 \subseteq G$ by
	\begin{align}
	L(x) \to g_0(x)L(x)h^{-1}(x)
	\end{align}
for $g_0 \in G_0$ and $h \in H$.
The Lagrangian is invariant under such transformations.

For future reference, we use the following index conventions:
	\begin{itemize}
	\item $I, J, \dotsc = 1, 2, \dotsc, 8$ for $\SO(8)$.
	\item $r, s, \dotsc = 9, 10, \dotsc, n+8$ for $\SO(n)$.
	\item $\bar I, \bar J,\dotsc = 1, 2, \dotsc, n+8$ for $\SO(8, n)$.
	\item $\cM, \cN, \dotsc$ for generators of $\SO(8, n)$. 	\end{itemize}	
Let the generators of $G$ be $\{t^\cM\} = \{t^{\bar I \bar J} \} = \{X^{IJ}, X^{rs}, Y^{Ir}\}$, where $Y^{Ir}$ are the noncompact generators.
Explicitly, the generators of the vector representation are given by
	\begin{align} 
	\tensor{(t^{\bar I \bar J})}{^{\bar K}_{\bar L}} = \eta^{\bar I \bar K} \delta^{\bar J}_{\bar L} - \eta^{\bar J \bar K} \delta^{\bar I}_{\bar L}
	\end{align}
where $\eta^{\bar I \bar J} = \diag( + + + + + + + + - \cdots)$ is the $\SO(8,n)$-invariant tensor.
These generators satisfy the typical $\SO(8, n)$ commutation relations,
	\begin{align} 
	[t^{\bar I \bar J}, t^{\bar K \bar L}] = 2\qty( \eta^{\bar I [\bar K} t^{ \bar L ] \bar J} - \eta^{\bar J [\bar K} t^{ \bar L ] \bar I} )
	\end{align}

The gauging of the supergravity is characterized by an embedding tensor $\Theta_{\cM \cN}$ (which has to satisfy various identities \cite{deWit:2003ja}) that determines which isometries are gauged, the coupling to the Chern-Simons fields, and additional terms in the supersymmetry transformations and action depending on the gauge coupling $g$.
We will look at the particular case in \cite{Berg:2001ty} where $n \geq 4$ and the gauged subgroup is the $G_0 = \SO(4) \times \SO(4)$ subset of the $\SO(8) \subset \SO(8, n)$.
The embedding tensor has the entries,\footnote{We use the conventions $\epsilon_{1234} = \epsilon_{5678} = 1$.}
	\begin{align} 
	\Theta_{\bar I \bar J, \bar K \bar L} = \begin{cases} \alpha \epsilon_{\bar I \bar J \bar K \bar L} & \text{if } \bar I, \bar J, \bar K, \bar L \in \{1, 2, 3, 4\} \\
		\epsilon_{\bar I \bar J \bar K \bar L} & \text{if } \bar I, \bar J, \bar K, \bar L \in \{5, 6, 7, 8\} \\ 0 & \text{otherwise} \end{cases}
	\end{align}
Note that the gauging depends on a real parameter $\alpha$.  
As discussed in  \cite{Berg:2001ty}, the maximally supersymmetric $\AdS_3$ vacuum has an isometry group,
	\begin{align}
	D^1(2,1;\alpha) \times D^1(2,1;\alpha)
	\end{align}
which corresponds to the family of ``large'' superconformal algebras of the dual SCFT.
In the following, we will consider the special case  $\alpha = 1$ for which $D^1(2, 1; 1) = \OSp(4|2)$ and the form of many quantities defined below are most compact. We expect that for generic values of $\alpha$ the qualitative behavior of the solutions will be similar.

From the embedding tensor, the $G_0$-covariant currents can be obtained,
	\begin{align} 
	L^{-1} (\partial_\mu + g \Theta_{\cM \cN} B_\mu^\cM t^\cN ) L = \frac{1}{2} \cQ^{IJ}_\mu X^{IJ} + \frac{1}{2} \cQ^{rs}_\mu X^{rs} + \cP^{Ir}_\mu Y^{Ir}
	\end{align}
It is convenient to define the $\tensor{\cV}{^\cM_\cA}$ tensors,
	\begin{align}  
	L^{-1} t^\cM L = \tensor{\cV}{^\cM_\cA} t^\cA = \frac{1}{2} \tensor{\cV}{^\cM_{IJ}} X^{IJ} + \frac{1}{2} \tensor{\cV}{^\cM_{rs}} X^{rs} + \tensor{\cV}{^\cM_{Ir}} Y^{Ir}
	\end{align}
and the $T$-tensor,
	\begin{align} 
	T_{\cA | \cB} = \Theta_{\cM \cN} \tensor{\cV}{^\cM_\cA} \tensor{\cV}{^\cN_\cB}
	\end{align}
The $T$-tensor is used to  construct the  tensors $A_{1, 2, 3}$ which will appear in the scalar potential and the supersymmetry transformations,
	\begin{align} \label{amatrix}
	A_1^{AB} &= - \frac{1}{48} \Gamma^{IJKL}_{AB} T_{IJ|KL} \nonumber \\
	A_2^{A\dot A r} &= - \frac{1}{12} \Gamma^{IJK}_{A\dot A} T_{IJ|Kr} \nonumber \\
	A_3^{\dot A r \dot B s} &=  \frac{1}{48} \delta^{rs} \Gamma^{IJKL}_{\dot A \dot B} T_{IJ|KL} + \frac{1}{2} \Gamma^{IJ}_{\dot A \dot B} T_{IJ|rs}
	\end{align}
where $A, B$ and $\dot A, \dot B$ are $\SO(8)$-spinor indices.
Our conventions for the $\SO(8)$ Gamma matrices are presented in appendix \ref{appendix-gamma}.

We take the spacetime signature $\eta^{ab} = \diag(+--)$ to be mostly negative.
The bosonic Lagrangian and scalar potential are
	\begin{align} \label{lagrangian}
	e^{-1} \cL_{\rm bos} &= - \frac{1}{4} R + \frac{1}{4} \cP_\mu^{Ir} \cP^{Ir\, \mu} + W - \frac{1}{4} e^{-1} \epsilon^{\mu\nu\rho} g \Theta_{\cM \cN} B_\mu^\cM \qty( \partial_\nu B_\rho^\cN + \frac{1}{3} g \Theta_{\cK \cL} \tensor{f}{^{\cN \cK}_{\cP}} B_\nu^\cL B_\rho^\cP ) \nonumber \\
	W &= \frac{1}{4} g^2 \qty( A^{AB}_1 A^{AB}_1 - \frac{1}{2} A^{A \dot A r}_2 A^{A \dot A r}_2 ) 
	\end{align}
The supersymmetry variations are
	\begin{align}
	\delta \chi^{\dot A r} &= \frac{1}{2} i \Gamma^I_{A\dot A} \gamma^\mu \epsilon^A \cP^{Ir}_\mu + g A^{A \dot A r}_2 \epsilon^A \nonumber \\
	\delta \psi^A_\mu &= \qty(\partial_\mu \epsilon^A + \frac{1}{4} \omega_\mu^{ab} \gamma_{ab} \epsilon^A + \frac{1}{4} \cQ^{IJ}_\mu \Gamma^{IJ}_{AB} \epsilon^B) + i g A^{AB}_1 \gamma_\mu \epsilon^B 
	\end{align}
The Einstein equation of motion is
	\begin{align} 
	R_{\mu\nu} - \cP^{Ir}_\mu \cP^{Ir}_\nu - 4 W g_{\mu\nu} = 0 
	\end{align}
and the gauge field equation of motion  is
	\begin{align} \label{gaugefieldeq}
	e \cP^{Ir\, \lambda} \Theta_{\cQ \cM} \tensor{\cV}{^\cM_{Ir}} = \epsilon^{\lambda \mu\nu} \qty( \Theta_{\cQ \cM} \partial_\mu B^\cM_\nu  + \frac{1}{6} g B^\cM_\mu B^\cK_\nu \qty( \Theta_{\cM \cN} \Theta_{\cK \cL} \tensor{f}{^{\cN \cL}_\cQ} + 2\Theta_{\cM \cN} \tensor{f}{^{\cL \cN}_\cK} \Theta_{\cL \cQ} ) )
	\end{align}
	 	
\subsection{The $n=4$ case}\label{sec21}
	
Let us focus on the case of four vector multiplets, i.e. $n = 4$.
The symmetries consist of a local $G_0 = \SO(4) \times \SO(4)$ and a global $\SO(n=4)$.
Thus, the scalar potential only depends on $8 \cdot 4 - 3 \cdot 6 = 14$ parameters out of the original $32$.
Moreover, we will only consider a further  consistent truncation outlined in \cite{Berg:2001ty} where the coset representative depends on eight of the fourteen scalars.
	\begin{align}
	L &= \mqty( \cos A & \sin A \cosh B & \sin A \sinh B \\ -\sin A & \cos A \cosh B & \cos A \sinh B \\ 0 & \sinh B & \cosh B) \nonumber \\
	A &= \diag(p_1, p_2, p_3, p_4)~, \qquad B = \diag(q_1, q_2, q_3 ,q_4)
	\end{align}
With this truncation, the scalar potential is\footnote{We correct a small typo in the potential given in \cite{Berg:2001ty}.}
	\begin{align}
	g^{-2} W &= 1 + \prod_{i=1}^4 \cosh q_i + \frac{1}{4}\sum_{i=1}^4 \sinh^2 q_i - \frac{1}{4} \sum_{i<j<k} (x_i^2 x_j^2 x_k^2 + y_i^2 y_j^2 y_k^2) - \frac{1}{2} \qty(\prod_{i=1}^4 x_i + \prod_{i=1}^4 y_i)^2 \nonumber \\
	x_i &= \cos p_i \sinh q_i ~, \qquad y_i = \sin p_i \sinh q_i
	\end{align}

The $\cQ_\mu$ and $\cP_\mu$ currents, excluding the $g \Theta_{\cM\cN} B^\cM_\mu \tensor{\cV}{^\cN_\cA}$ term, are
	\begin{align}
	\cQ^{IJ}_\mu &=  \qty(\smqty{ 0&0&0&0& \cosh q_1 \partial_\mu p_1 &0&0&0 \\
						0&0&0&0& 0& \cosh q_2 \partial_\mu p_2 &0&0 \\
						0&0&0&0& 0&0& \cosh q_3 \partial_\mu p_3 &0 \\
						0&0&0&0& 0&0&0& \cosh q_4 \partial _\mu p_4 \\
						-\cosh q_1 \partial_\mu p_1 &0&0&0 & 0&0&0&0 \\
						0& -\cosh q_2 \partial_\mu p_2 &0&0 & 0&0&0&0 \\
						0&0& -\cosh q_3 \partial_\mu p_3 &0 & 0&0&0&0 \\
						0&0&0& -\cosh q_4 \partial _\mu p_4 & 0&0&0&0 })_{IJ}
	\nonumber \\
	\cQ^{rs}_\mu &= 0 \nonumber \\
	\cP^{Ir}_\mu &= \qty(\smqty{ \sinh q_1 \partial_\mu p_1 & 0 & 0 & 0 \\
						  0 & \sinh q_2 \partial_\mu p_2 & 0 & 0 \\
						  0 & 0 & \sinh q_3 \partial_\mu p_3 & 0 \\
						  0 & 0 & 0 & \sinh q_4 \partial_\mu p_4 \\
						  \partial_\mu q_1 & 0 & 0 & 0 \\
						  0 & \partial_\mu q_2 & 0 & 0 \\
						  0 & 0 & \partial_\mu q_3 & 0 \\
						  0 & 0 & 0 & \partial_\mu q_4 })_{Ir}
	\end{align}
Using these matrices, we can check that the combination $\cP^{Ir}_\mu \tensor{\cV}{^{JK}_{Ir}} $ vanishes whenever the indices $J, K \in \{1, 2, 3, 4\}$ or $J, K \in \{5, 6, 7, 8\}$.
This implies that there is no source for $B^\cM_\mu$ in the gauge field equation of motion (\ref{gaugefieldeq}), so it is consistent to set $B^\cM_\mu = 0$.
We will make this choice from now on.

\subsection{Truncations and supersymmetric $\AdS_3$ vacua}
	
In order to make our analysis more tractable, we make further truncations to reduce the number of independent scalar fields.\footnote{These truncations are consistent due to internal rotational symmetry as shown in \cite{Berg:2001ty}.} Below we consider three truncations, which together explore the  $\AdS_3$ vacua with $\cN = (4,4)$ and $\cN = (1,1)$ supersymmetry.

\subsubsection{Truncation 1}\label{trunc1}

The first truncation is given by calling $q_1 = q$, $p_1 = p$ and setting all remaining $q_i = p_i = 0$ for $i = 2,3,4$.
The scalar potential is
	\begin{align}
	W = \frac{g^2}{2} \cosh^2 \frac{q}{2} \, (3 + \cosh q)
	\end{align}
The $\cN=(4,4)$ vacuum is given  by setting $q=0$ and the vacuum potential is $W_0 = 2 g^2$.

\subsubsection{Truncation 2}\label{trunc2}

The second truncation is given by setting all the $q$s and $p$s equal, i.e.~$q_i=q$, $p_i=p$ for $i=1,2,3,4$.  
The scalar potential is
	\begin{align}
	W = \frac{g^2}{8192} \Big( 8103 &+ 6856 \cosh 2 q + 1452 \cosh 4 q - 8 \cosh 6 q - 19 \cosh 8 q \nonumber \\
		& - 768 (3 + \cosh 2 q) \cos 4 p \sinh^6 q - 128 \cos 8 p \sinh^8 q \Big)
	\end{align}
The $\cN=(4,4)$ vacuum is given by $q=0$ as before, and $\cN=(1,1)$ vacua are given by $q = \pm \sinh^{-1} \sqrt{2}$ and $p = \pi(\mathbb{Z}/2 + 1/4)$ which have a vacuum potential of $W_0 = 8 g^2$.

\subsubsection{Truncation 3}\label{trunc3}

The third truncation is given by setting the first three $q$s and $p$s equal, i.e.~$q_i = q$, $p_i = p$ for $i = 1, 2, 3$, and setting the remaining $q_4 = p_4 = 0$.
The scalar potential is 
	\begin{align}
	W = \frac{g^2}{1024} \Big(690 &+ 768 \cosh q + 309 \cosh 2 q +  256 \cosh 3 q \nonumber \\
		&+ 30 \cosh 4 q - 5 \cosh 6q - 96 \cos 4 p \sinh^6 q \Big)
	\end{align}
The $\cN=(4,4)$ vacuum is given by $q=0$ as before, and $\cN=(1,1)$ vacua are given by $q= \pm \sinh^{-1} \sqrt{ 2+2\sqrt{2}}$ and $p = \pi(\mathbb{Z}/2 + 1/4)$ which have a vacuum potential of $W_0 = 2(1+\sqrt{2})^2g^2$.

\section{Janus flow equations }\label{sec3}
In this section, we present the equations of motion and supersymmetry variations for a Janus ansatz where the three-dimensional metric is written as an $\AdS_2$ slicing and the scalar fields only depend on the slicing coordinate. 
We will also set the Chern-Simons gauge $B_\mu^{\cal M}$  fields to zero, which is consistent as argued in  section \ref{sec21}. Hence, the Janus ansatz is given by
	\begin{align} \label{janusan}
	\dd{s^2} &= e^{2B(u)} \qty(\frac{ \dd{t^2} - \dd{z^2} }{z^2}) -\dd{u}^2~, \qquad  B^\cM_\mu  = 0 \nonumber \\
	q_i &= q_i(u)~, \qquad  p_i = p_i(u)
	\end{align}
The Ricci tensor has the non-zero components,
	\begin{align}
	R_{tt} = - R_{zz} &= z^{-2} \qty( 1 + e^{2B}(2 B'^2  + B'')  ) \nonumber \\
	R_{uu} &= -2 (B'^2  + B'') 
	\end{align}
The prime $'$ denotes a derivative with respect to the slicing coordinate $u$.
The gravitino supersymmetry  variation  $\delta \psi^A_\mu = 0$ is
	\begin{align}\label{gravvar}
	0 &= \partial_t \epsilon + \frac{1}{2z} \gamma_0 \qty(\gamma_1  - B' e^{B} \gamma_2 + 2 i g e^B A_1) \epsilon \nonumber \\
	0 &= \partial_z \epsilon + \frac{1}{2z} \gamma_1 \qty( - B' e^{B} \gamma_2 + 2 i g e^B A_1 ) \epsilon \nonumber \\
	0 &= \partial_u \epsilon + \frac{1}{4} \cQ_u^{IJ} \Gamma^{IJ} \epsilon + i g  \gamma_2 A_1 \epsilon 
	\end{align}
where we have suppressed the $\SO(8)$-spinor indices of $\epsilon^A$ and $A_1^{AB}$. There are two  integrability conditions  which can be derived from the gravitino variations (\ref{gravvar}) in the $t,z$ and $z,u$ directions respectively
	\begin{align}\label{integra}
	0 &= \qty(1 -(2 g e^B A_1)^2 + (B' e^{B})^2 )\epsilon \nonumber \\
	0 &= 2ig e^B \qty( A_1' - \frac{1}{4} [ A_1, \cQ_u^{IJ} \Gamma^{IJ}] ) \epsilon + \qty( - \dv{u} (B' e^{B}) + (2 g e^B A_1)^2 e^{-B} ) \gamma_2 \epsilon 
	\end{align}
We can use the first integrability condition to express the second one as
	\begin{align}\label{integrab}
	2ig   \qty( A_1' - \frac{1}{4} [ A_1, \cQ_u^{IJ} \Gamma^{IJ}] ) \epsilon + \qty( -  B'' + e^{-2B} ) \gamma_2 \epsilon=0 
	\end{align}
The spin-${1\over 2}$  variation $\smash{\delta \chi^{\dot A } = 0}$ is
	\begin{align} \label{integrac}
	\qty(- \frac{i}{2} \Gamma^I \cP_u^{I r} \gamma_2 + g A_2^{r} )_{A\dot A} \epsilon^A = 0~, \quad \quad r=9,10,\dotsc, 8+n
	\end{align}

\subsection{Eigenvectors of $A_1$}
	
It follows from (\ref{amatrix})  that  $A_1$ is a $8\times 8$ matrix which has eigenvectors 
	\begin{align}
	A_1^{AB} n_\pm^{(i) B} = \pm w_i   n_{\pm}^{(i) A}~, \quad \quad i=1,2,3,4
	\end{align}
The eigenvalues $w_i$ determine whether a supersymmetric $\AdS_3$ vacuum can exist.  In the following we denote the positive supersymmetric eigenvalue $w$, which can be determined as follows: for the $\AdS_2$-sliced metric given in (\ref{janusan}), the $\AdS_3$ vacuum solution with potential $W_0$ is given by 
	\begin{align} \label{vacuumb}
	B_{\rm vac}(u) =\ln {\cosh (\sqrt{2 W_0} u) \over \sqrt{2 W_0}}
	\end{align}
which satisfies 
	\begin{align}
	B_{\rm vac}'^2+ e^{-2B_{\rm vac}} - 2 W_0 = 0
	\end{align}
When we expand the spinors $\epsilon^A$ in terms of the eigenvectors of $A_1$, the first equation in (\ref{integra}) implies for the spinor component associated with the eigenvalue $w$ that
	\begin{align}\label{brimeb}
	B'^2 + e^{-2B}- 4 g^2 w^2=0
	\end{align}
For the $\AdS_3$ vacuum solution (\ref{vacuumb}), this condition relates the eigenvalue evaluated at the vacuum  $w_{\rm vac}$ to the potential $W_0$ via
	\begin{align}\label{susyvac}
	w^2_{\rm vac} = {W_0\over 2 g^2} 
	\end{align}

As discussed in section \ref{sec4} for truncation 1, $A_1$ has eight eigenvectors $n_\pm^{(i)}$ for $i=1,2,3,4$ all with the same with eigenvalue $\pm w$ that satisfy the supersymmetry condition (\ref{susyvac}) for the $\AdS_3$ vacuum with $q=0$. Hence, this vacuum preserves $\cN=(4,4)$ supersymmetry. On the other hand for truncations 2 and 3, there are only two eigenvectors $n_\pm$ with an eigenvalue $\pm w$ that satisfy (\ref{susyvac})  for the $\AdS_3$ vacua with non-trivial values for the scalars. Consequently, these vacua only preserve $\cN=(1,1)$ supersymmetry.

For the RG-flow solutions which interpolate between the different vacua, we expand the spinors in the basis of the eigenspinors that correspond to the supersymmetric vacuum when the scalars take their vacuum values. 
This implies that (\ref{brimeb}) can  be solved for $B'$,
	\begin{align}\label{bprime}
	B'&= \pm \sqrt{4 g^2 w^2 -e^{-2B}} \nonumber\\
	&= \pm 2 g w \gamma
	\end{align}
where we defined  the convenient combinations,
	\begin{align}\label{gamdef}
	\gamma(u) = \sqrt{1- {e^{-2B}\over 4 g^2 w^2}}~, \quad \quad \sqrt{1-\gamma^2(u)}= {e^{-B}\over 2g w}
	\end{align}
	which will be useful later on.
The two signs in (\ref{bprime}) are two branches of solutions which for Janus solutions will be patched together---the numerical evolution usually breaks down at $B'=0$ and this is the location where the two branches will be glued together.

\subsection{$\AdS_2$ Killing spinors}

The Killing spinors for  a  unit radius $\AdS_2$ with metric,
\begin{align}
\dd{s_{\AdS_2}^2}={ \dd{t^2}-\dd{z^2} \over z^2 }
\end{align}
 satisfy the following equation,
	\begin{align}
	D_\mu \zeta_\eta  = i {\eta\over 2}\gamma_\mu \zeta_\eta~, \quad \quad \mu=t, z
	\end{align} 
with $\eta=\pm 1$. 
The covariant derivatives on $\AdS_2$  take the form,
	\begin{align}
	D_t= \partial_t \epsilon + \frac{1}{2z} \gamma_0\gamma_1~,\quad \quad D_z= \partial_z
	\end{align}
Since the general spinor in $\AdS_2$ is a two-component spinor, the $\zeta_\pm$ form a basis of two-component spinors. 
Since spinors in three dimensions are also two-component spinors, the $\zeta_\pm$ are also a basis of the spinors in three dimensions. 
Note that $\gamma_2 =  i \gamma_\#$ where $ \gamma_\#^2=1$ and 
	\begin{align}
	i \gamma_2 \zeta_\eta=\zeta_{-\eta}~, \quad \quad \eta=\pm 1
	\end{align}

The general ansatz for $\epsilon^A$ is given by
	\begin{align} \label{spinor-ansatz}
	\epsilon^A  = \sum_i \big(f_+^{(i)} n_+^{(i) A} + f_-^{(i)} n_-^{(i) A}\big) \zeta_{+}+\big(g_+^{(i)} n_+^{(i) A} + g_-^{(i)} n_-^{(i) A}\big) \zeta_{-}
	\end{align}
For truncation 1 we have $i=1,2,3,4$, which label  four eigenvectors of $A_1$,  whereas in truncations 2 and 3 the index $i$ is dropped.

\subsection{First projector}
With this ansatz for the spinors $\epsilon^A$, the first two components of the gravitino variation,
\begin{align}
	0 &= \partial_t \epsilon + \frac{1}{2z} \gamma_0 \qty(\gamma_1  - B' e^{B} \gamma_2 + 2 i g e^B A_1) \epsilon \nonumber \\
	0 &= \partial_z \epsilon + \frac{1}{2z} \gamma_1 \qty( - B' e^{B} \gamma_2 + 2 i g e^B A_1 ) \epsilon
\end{align}
can be expressed as follows by using the properties of the $\AdS_2$ Killing spinors,
	\begin{align}
	0&= i \qty{ \big(f_+^{(i)} n_+^{(i) A} + f_-^{(i)} n_-^{(i) A}\big) \zeta_{+} - \big(g_+^{(i)} n_+^{(i) A} + g_-^{(i)} n_-^{(i) A}\big)  \zeta_{-} } \\
	&\quad  + i B'e^{-B} i \gamma_2 \Big\{ \big(f_+^{(i)} n_+^{(i) A} + f_-^{(i)} n_-^{(i) A}\big) \zeta_{+}+\big(g_+^{(i)} n_+^{(i) A} + g_-^{(i)} n_-^{(i) A}\big) \zeta_{-}\Big\}\nonumber\\
	&\quad + 2i g w e^{B} \Big\{ \big(f_+^{(i)} n_+^{(i) A} - f_-^{(i)} n_-^{(i) A}\big) \zeta_{+}+\big(g_+^{(i)} n_+^{(i) A} - g_-^{(i)} n_-^{(i) A}\big) \zeta_{-}\Big\}
	\end{align}
Using $i\gamma_2 \zeta_{\eta }= \zeta_{-\eta}$ and the linear independence of the $n_{\pm}^{(i)}$  and $\zeta_{\pm}$, one obtains a set of equations,
	\begin{align}
	f_+ + B' e^B g_+ + 2 g w e^B f_+&=0\nonumber \\
	-g_+ + B' e^B f_+ + 2 g w e^B g_+&=0\nonumber \\
	f_- + B' e^B g_- - 2 g w e^B f_-&=0\nonumber \\
	-g_- + B' e^B f_- - 2 g w e^B g_-&=0
	\end{align}
which are consistent  if the integrability condition (\ref{brimeb}) holds. 
In terms of the $\gamma(u)$ defined in (\ref{gamdef}) we have
	\begin{align}\label{freplace}
	f_+ = {\sqrt{1-\gamma^2} -1\over \gamma}g_+~, \qquad  f_- = {\sqrt{1-\gamma^2} +1\over \gamma}g_-
	\end{align}

\subsection{Second projector}

The spin-${1\over 2}$  variation  (\ref{integrac}) can be rewritten in the following form
	\begin{align} \label{mmatrix-flow}
	\left[- \frac{1}{2g} \Big((A_2^{r})^{T} \Big)^{-1}  \Big(\Gamma^I \cP_u^{I r} \Big)^T i\gamma_2 + 1 \right]^{AB}  \epsilon^B= 0~, \quad \quad r=9,10,11,12
	\end{align}
Since $\cP^{Ir}_u$ contains the first derivatives of the scalar fields, the flow equations for the scalars  can be derived  from the condition of  vanishing of this supersymmetry variation.
The projectors for $r=9,10,11,12$ take the form
	\begin{align}\label{proj1}
	\Big(M^{AB} i \gamma_2 + \delta^{AB} \Big) \epsilon^B = 0
	\end{align}
For consistency, the matrix must  satisify  $M^{AB} M^{BC} = \delta^{AC}$. 
Plugging in the ansatz (\ref{spinor-ansatz}) for the spinors $\epsilon^A$, we get
	\begin{align}
	0&= \big(f_+^{(i)} n_+^{(i) A} + f_-^{(i)} n_-^{(i) A}\big) \zeta_{+}+\big(g_+^{(i)} n_+^{(i) A} + g_-^{(i)} n_-^{(i) A}\big) \zeta_{-}\nonumber \\
	&\quad + M^{AB} i\gamma_2  \Big\{  \big(f_+^{(i)} n_+^{(i) B} + f_-^{(i)} n_-^{(i) B}\big) \zeta_{+}+\big(g_+^{(i)} n_+^{(i) B} + g_-^{(i)} n_-^{(i) B}\big)\zeta_{-}\Big\}
	\end{align}
Using the fact that the eigenvectors can be orthonormalized,
	\begin{align}
	n_+^{(i) A} n_+^{(j) A} =\delta^{ij}~, \qquad  n_-^{(i) A} n_-^{(j) A} =\delta^{ij}~, \qquad  n_+^{(i) A} n_-^{(j) A} =0
	\end{align}
and projecting onto $n_{\pm}^{(i)}$ gives
	\begin{align}\label{fgcon}
	f_+ + M_{++} g_+ + M_{+-} g_-&=0\nonumber \\
	g_+ + M_{++} f_+ + M_{+-} f_-&=0\nonumber \\
	f_- + M_{+-} g_+ + M_{--} g_-&=0\nonumber \\
	g_+ + M_{+-} f_+ + M_{--} f_-&=0 
	\end{align}
where we define
	\begin{align}
	M_{++}= n_{+}^A   M^{AB} n_+^B~, \qquad  M_{--}= n_{-}^A   M^{AB} n_-^B~, \qquad  M_{+-}= M_{-+}= n_{+}^A   M^{AB} n_-^B
	\end{align}
If there is more than one $n_\pm$ (as in truncation 1) the $M_{\pm\pm}, M_{\pm\mp}$ have to take the same form for all $n^{(i)}_\pm$, which is a consistency condition. 
Using (\ref{freplace}) it can be shown that  equations  (\ref{fgcon}) can only\footnote{We can also have $M_{+-} = M_{-+} = - \sqrt{1-\gamma^2}$, which gives a similar solution. For example, in section \ref{sec4-1} for truncation 1, this sends $p(u) \to p(-u)$. This resolves an ambiguity in the definition of our eigenvectors, as we can freely send $n_+ \to - n_+$ or $n_- \to -n_-$.} be satisfied if we have
	\begin{align}\label{mmatrix}
	M_{++}= \gamma~, \qquad M_{--} =-\gamma~, \qquad M_{+-}=M_{-+}=\sqrt{1-\gamma^2}
	\end{align}
The relations (\ref{mmatrix}) for the matrix $M^{AB}$ given in (\ref{mmatrix-flow}) provide first-order flow equations for the scalar fields.  
Note that the second integrability condition for the gravitino variation in (\ref{integra}) is also  the form of (\ref{proj1}). For all the solutions which we find, this condition is automatically satisfied and does not constrain the flow further.

\section{Janus and RG-flow solutions }\label{sec4}

In this section we obtain the flow equations and solve them.
Only for truncation 1 are we able to solve the system analytically. For truncations 2 and 3 we solve the resulting flow equations numerically.

\subsection{Truncation 1} \label{sec4-1}
\begin{figure}[htbp]
\begin{center}
{\centerline{\includegraphics[width=10cm]{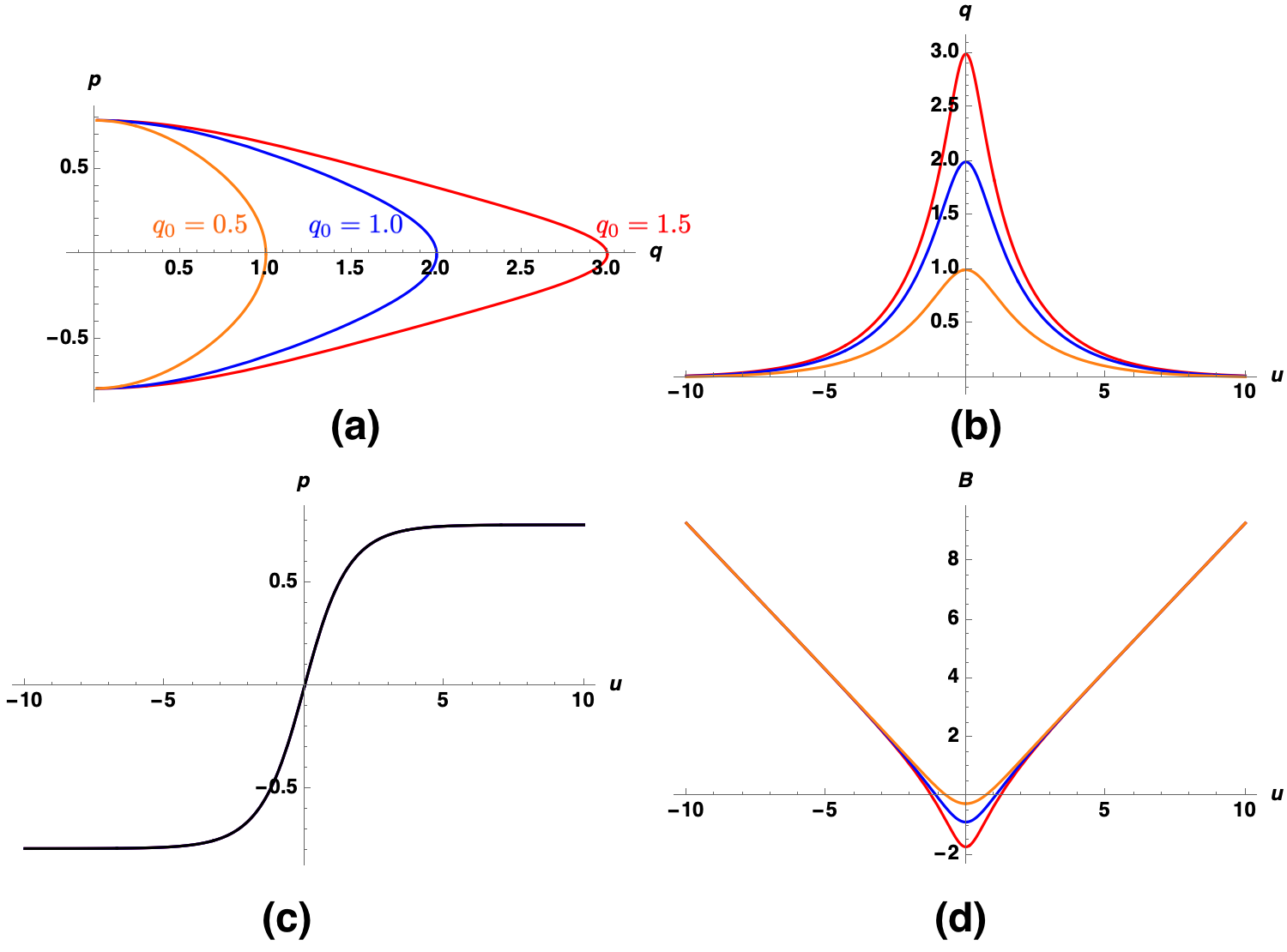}}}
\caption{ (a) $pq$ parametric plot, (b) plot of $q$ , (c) plot of $p$, (d) plot of the metric function $B$ as functions of the $\AdS_2$ slicing coordinate $u$ for truncation 1.  The colors denote three different values for $q_0$. $p_0=0$ for all three plots. The behavior of $p$ is the same for all three examples.}
\label{fig:fig1}
\end{center}
\end{figure}

For the truncation to a single scalar described in section \ref{trunc1}, the matrix $A_1$ takes the form,
	\begin{align}  
	A_1 &= \tiny{\left( \begin{array}{cccccccc}
		 0 & 0 & 0 & \cos p\cosh^2 {q\over 2} & 0 & 0 & \sin p\cosh^2 {q\over 2} & 0 \\
		 0 & 0 & -\cos p\cosh^2 {q\over 2} & 0 & 0 & 0 & 0 & \sin p\cosh^2 {q\over 2} \\
		 0 & -\cos p\cosh^2 {q\over 2} & 0 & 0 & -\sin p\cosh^2 {q\over 2} & 0 & 0 & 0 \\
		 \cos p\cosh^2 {q\over 2} & 0 & 0 & 0 & 0 &- \sin p\cosh^2 {q\over 2} & 0 & 0 \\
		 0 & 0 &- \sin p\cosh^2 {q\over 2} & 0 & 0 & 0 & 0 & -\cos p\cosh^2 {q\over 2} \\
		 0 & 0 & 0 &- \sin p\cosh^2 {q\over 2} & 0 & 0 & \cos p\cosh^2 {q\over 2} & 0 \\
		 \sin p\cosh^2 {q\over 2} & 0 & 0 & 0 & 0 & \cos p\cosh^2 {q\over 2} & 0 & 0 \\
		 0 & \sin p\cosh^2 {q\over 2} & 0 & 0 & -\cos p\cosh^2 {q\over 2} & 0 & 0 & 0 \\
	\end{array} \right )}
	\end{align}
We have four pairs of eigenvectors $n_\pm^{(i)}$ for $i=1,2,3,4$ with the same  eigenvalues $\pm w$, where 
	\begin{align}
	w=\cosh^2 {q\over 2}
	\end{align}
so the supersymmetry condition (\ref{susyvac}) is satisfied for the vacuum where $q=0$.
The pairs of eigenvectors are
	\begin{align}
	n^{(1)}_+ &= \qty{\mqty{ \frac{1}{\sqrt{2}}, 0, 0, \frac{\cos p}{\sqrt{2}} , 0, 0, \frac{ \sin p}{\sqrt{2}}, 0}}~, &
	n^{(1)}_- &= \qty{\mqty{ 0, 0, 0, -\frac{\sin p}{\sqrt{2}} , 0, -\frac{1}{\sqrt{2}}, \frac{\cos p}{\sqrt{2}} , 0}} \nonumber  \\
	n^{(2)}_+ &= \qty{\mqty{ 0, \frac{1}{\sqrt{2}}, -\frac{\cos p}{\sqrt{2}} , 0, 0, 0, 0, \frac{\sin p}{\sqrt{2}} }}~, &
	n^{(2)}_- &= \qty{\mqty{ 0, 0, \frac{\sin p}{\sqrt{2}} , 0, \frac{1}{\sqrt{2}} , 0, 0, \frac{\cos p}{\sqrt{2}} }} \nonumber \\
	n^{(3)}_+ &= \qty{\mqty{ 0, 0, -\frac{\sin p}{\sqrt{2}} , 0, \frac{1}{\sqrt{2}} , 0, 0, -\frac{\cos p}{\sqrt{2}} }} ~, &
	n^{(3)}_- &= \qty{\mqty{ 0, -\frac{1}{\sqrt{2}}, -\frac{\cos p}{\sqrt{2}} , 0, 0, 0, 0, \frac{\sin p}{\sqrt{2}}}} \nonumber \\
	n^{(4)}_+ &= \qty{\mqty{ 0, 0, 0, -\frac{\sin p}{\sqrt{2}} , 0, \frac{1}{\sqrt{2}}, \frac{\cos p}{\sqrt{2}} , 0}}~, &
	n^{(4)}_- &= \qty{\mqty{ \frac{1}{\sqrt{2}}, 0, 0, -\frac{\cos p}{\sqrt{2}} , 0, 0, -\frac{\sin p}{\sqrt{2}} , 0}}
	\end{align}
Given the eigenvectors, we can compute the $M_{++}$ and $M_{+-}$ matrix elements for the matrix in (\ref{mmatrix-flow}) for any pair of $n^{(i)}_\pm$. Note that for this truncation, only the flow equation for index $r=9$ is nontrivial while the others are identically vanishing.
Then (\ref{mmatrix}) gives us flow equations for the scalars $q$ and $p$.
The remaining flow equation for the metric factor $B$ comes from (\ref{bprime}).
The flow equations are
	\begin{align}
	q'= -g \gamma \sinh q ~, \qquad p'= g \sqrt{1-\gamma^2} ~, \qquad B'= \pm 2 g \gamma  \cosh^2 {q\over 2} 	
	\end{align}
which solve the equations of motion.
We can solve for a flow where $p(0) = p_0$, $q(0) = q_0$ are arbitrary and $B'(0) = 0$, which is equivalent to $\gamma(0) = 0$.
This first-order system can be rewritten using the function $\gamma$ in lieu of $B$, in which case the third equation above is replaced with
	\begin{align}
	\gamma' =  2 g (1 - \gamma^2) 	
	\end{align}
The solution is
	\begin{align}
	\gamma(u) &= \tanh(2g u) \nonumber \\
	\tanh \frac{q(u)}{2} &= \sqrt{\sech(2gu)} \tanh \frac{q_0}{2} \nonumber \\
	 \qquad \tan[p(u) - p_0] &= \tanh(gu)
	\end{align}	
which gives the metric factor
	\begin{align}
	e^{B(u)} = \frac{\cosh(2gu)}{2g} \sech^2 \frac{q(u)}{2} 
	\end{align}
This is a Janus solution which approaches the $\cN = (4, 4)$ vacuum at the two endpoints $u \to \pm \infty$.
We note that the flow equation is the same for each pair of eigenvectors $n^{(i)}_\pm$ and hence the solution preserves eight of the sixteen supersymmetries of the $\cN=(4,4)$ vacuum. We present plots for three choices of the parameters $q_0=0.5, 1.0,1.5$ in figure \ref{fig:fig1} and set $p_0=0$ for all three. The qualitative behavior is very similar for all three choices and corresponds to a Janus interface which interpolates between different values of $p(u)$ as $u\to \pm \infty$. 

\subsection{Truncation 2}
Recall that the truncation presented in section \ref{trunc2} sets all the $q_i$ equal and $p_i$ equal. 
The matrix $A_1$ takes the form,
	\begin{align} A_1 &= \tiny{\left( \begin{array}{cccccccc}
		 0 & 0 & 0 & a_1+a_3 & 0 & 0 & 0 & 0 \\
		 0 & -2a_3 & -a_1+a_3 & 0 & a_2 & 0 & 0 & a_2 \\
		 0 & -a_1+a_3 & -2a_3 & 0 & -a_2 & 0 & 0 & -a_2 \\
		 a_1+a_3 & 0 & 0 & 0 & 0 & 0 & 0 & 0 \\
		 0 & a_2 & -a_2 & 0 & 2a_3 & 0 & 0 & -a_1+a_3 \\
		 0 & 0 & 0 & 0 & 0 & 0 & a_1+a_3 & 0 \\
		 0 & 0 & 0 & 0 & 0 & a_1+a_3 & 0 & 0 \\
		 0 & a_2 & -a_2 & 0 & -a_1+a_3 & 0 & 0 & 2a_3 \\
	\end{array} \right) }
	\end{align}
where we define, for this truncation,
	\begin{align}
	a_1&= {1\over 8} (3+ \cos 4p) (1+ \cosh^4 q) \nonumber \\
	a_2&={1\over 8} (3+ \cosh 2q) \cosh q \sin 4p \nonumber \\
	a_3&= 2 \cos^2 p \sin^2 p \cosh^2q
	\end{align}
The eigenvectors and eigenvalues can be obtained by considering  the matrix $(A_1)^2$ first.
There are two eigenvalues, the first is $(a_1 + a_3)^2$ which is six-fold degenerate but does not satisfy the condition (\ref{susyvac}) for the $\cN=(1,1)$ vacuum.
The second eigenvalue is $4 a_2^2 + (a_1 - 3 a_3)^2$ which is two-fold degenerate and does satisfy (\ref{susyvac}) for the $\cN=(1,1)$ vacua. 
The corresponding eigenvectors of $(A_1)^2$ take the form
	\begin{align} \label{v1v2}
	v_1 = \left\{\mqty{0,\frac{1}{\sqrt{2}},-\frac{1}{\sqrt{2}},0,0,0,0,0}\right\}~, \qquad v_2 = \left\{\mqty{0,0,0,0,\frac{1}{\sqrt{2}},0,0,\frac{1}{\sqrt{2}}}\right\}
	\end{align}
Let 
	\begin{align}
	w = \sqrt{4 a_2^2 + (a_1 - 3 a_3)^2}
	\end{align}
be the positive eigenvalue of $A_1$.
One can reduce the $A_1$ matrix on the subspace spanned by $v_1, v_2$ and find that the (not yet normalized) eigenvectors $v_\pm$ of $A_1$ with eigenvalue $\pm w$ are given by
	\begin{align} 
	v_\pm &= 2a_2 v_1 + \qty(-a_1  + 3 a_3 \pm w) v_2
	\end{align}

The flow equations take the form of a first-order system of ordinary differential equations for the functions $p(u)$, $q(u)$, and $B(u)$. 
These equations do not take a simple form and are too unwieldy to be presented here.  Using Mathemtica, we have checked that the flow equations  imply that the equations of motion are satisfied  as well as the second integrability condition of the gravitino variation (\ref{integrab}).

\begin{figure}[htbp]
\begin{center}
{\centerline{\includegraphics[width=10cm]{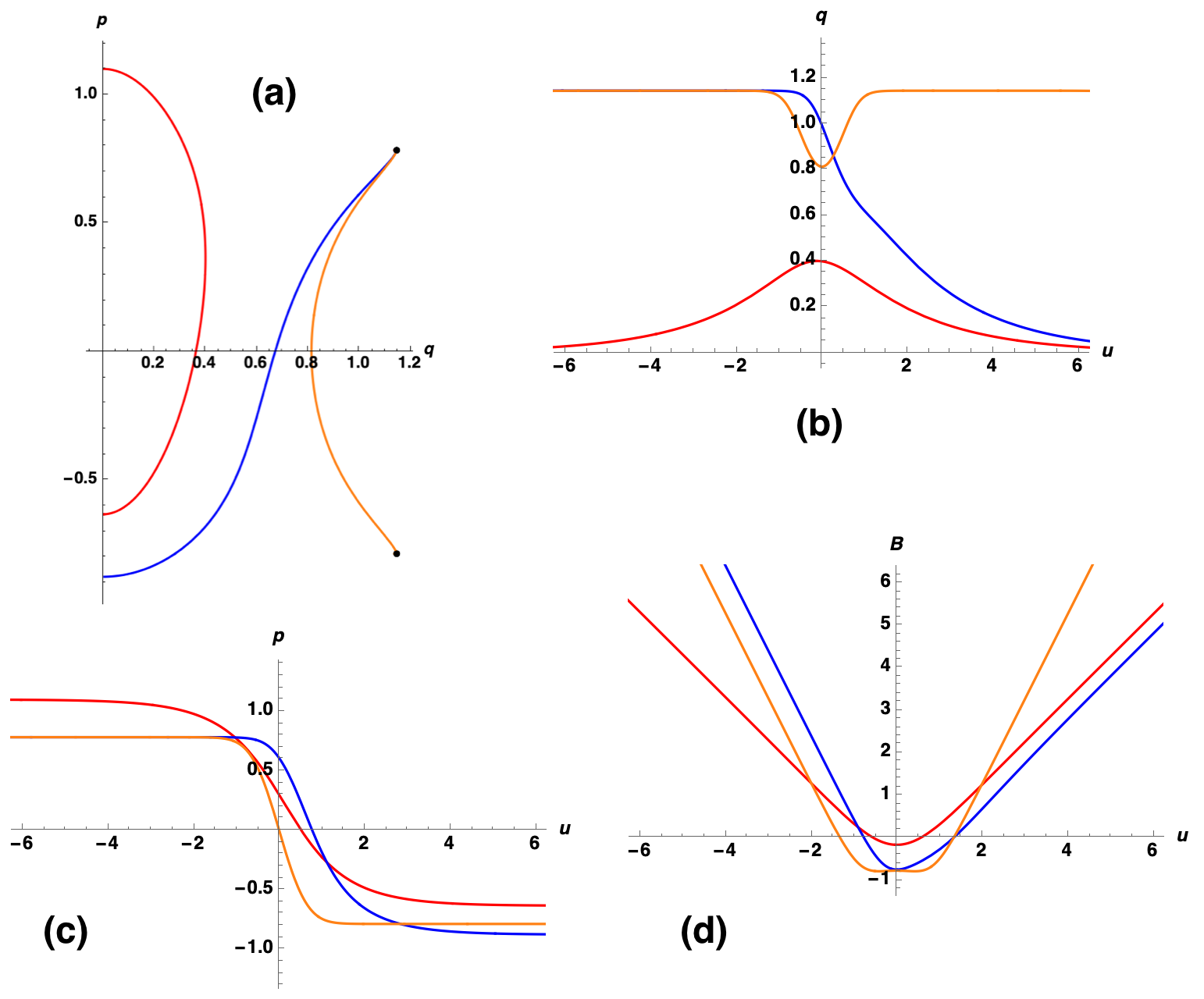}}}
\caption{(a) $pq$  parametric plot, (b) plot of $q$ , (c) plot of $p$, (d) plot of the metric function $B$ as functions of the $\AdS_2$ slicing coordinate $u$.}
\label{fig:fig2}
\end{center}
\end{figure}

The flow equations can be numerically integrated.\footnote{We use the method described in \cite{Bobev:2013yra}: we choose  the location $p(0),q(0)$ of a turning point where $B'=0$ and use the flow equations to determine $p'(0),q'(0)$. These values  provide  the  initial conditions for the second order equations of motion to give $p(u), q(u)$ and $B(u)$.}   In figure \ref{fig:fig2} we present some examples for the numerical solutions of the flow equations.
By fine-tuning initial conditions we can produce    flows that (i) look like the Janus solutions in truncation 1 plotted, in red in  figure \ref{fig:fig2}, (ii) connect $\cN=(4,4)$ and $\cN=(1,1)$ vacua,  plotted in blue in  figure \ref{fig:fig2}, and (iii) connect two $\cN=(1,1)$ vacua which are related by flipping signs of $p$, plotted in orange in  figure \ref{fig:fig2}. In the $pq$ parametric  plot  in  figure \ref{fig:fig2}(a), the locations of the $\cN=(1,1)$ vacua  $p=\pm {\pi\over 4}$, $q= \sinh^{-1}\sqrt{2}$ are denoted by black dots. Note that for the $\cN=(4,4)$ vacuum where $q=0$, the asympototic value  $p$ is a modulus which can take any value, hence  the flow towards the $\cN=(4,4)$ can end at any point on the $p$-axis at $q=0$.

In the numerical integration, the $\cN=(1,1)$ points are repulsive and require fine-tuning in order to obtain flows that approach these points.
This can be explained as follows.
The choice $g = 1/2$ sets the $\cN=(4,4)$ vacuum potential to $W_0 = 1/2$ and the $\AdS_3$ length scale to unity.
For the $\cN=(1,1)$ point, the vacuum potential becomes $W_0 = 2$ and the $\AdS_3$ length scale is $L = 1/2$.
Taking the linear expansion around the $\cN=(1,1)$ point,
	\begin{align}
	p(u) &= \frac{\pi}{4} + \delta p(u) + \order{\delta p^2} \nonumber \\
	q(u) &= \sinh^{-1} \sqrt{2} + \delta q(u) + \order{\delta q^2} 
	\end{align}
the mass-squares for the $\delta p$ and $\delta q$ fluctuations are
	\begin{align}
	m_p^2L^2 &= \frac{5}{4}~, \qquad m_q^2L^2 = \frac{21}{4}
	\end{align}
Using $\Delta (\Delta - 2) = m^2 L^2$, the scaling dimensions of the corresponding dual operators are
	\begin{align} \label{scalingdims2}
	\Delta_p &= \frac{5}{2}~, \qquad \Delta_q = \frac{7}{2}
	\end{align}
In our $\AdS$-sliced coordinates, the boundary is given by the two $\AdS_2$ components at $u = \pm\infty$, which are joined together at the $z = 0$ interface. 
The coordinates $(z, u)$ can be mapped to Fefferman-Graham coordinates $(\rho, x)$ where the boundary is located at $\rho = 0$.\footnote{Recall that the $\AdS_3$ metric in Poincar\'e coordinates,
	\begin{align*}
	\dd{s^2} = \frac{- \dd{\rho^2} + \dd{t^2} - \dd{x^2}}{\rho^2}
	\end{align*}
is related to an $\AdS_2$-sliced metric by the coordinate change,
	\begin{align*}
	z &= \sqrt{x^2 + \rho^2} & \sinh u = x / \rho
	\end{align*}
}
Let us consider the boundary at $u \to + \infty$.
In the $(\rho, x)$ coordinates, the metric factor $B$ has the expansion $e^B = \rho^{-1} + \order{\rho^0}$ near the boundary.
But in the $(z, u)$ coordinates, from (\ref{vacuumb}) the expansion near the boundary is $B = u/L + \cdots$.
Therefore, the asymptotic form of the coordinate change $(\rho, x) \mapsto (z, u)$ takes the form,
	\begin{align}
	e^u &= \rho^{-L} + \cdots
	\end{align}
The linearized flow equations around the $\cN=(1,1)$ point are
	\begin{align}
	\delta p' &= -5 \delta p + \cdots \nonumber \\
	\delta q' &= 3 \delta q + \cdots
	\end{align}
which are solved by $\delta p \sim C_p e^{-5u}$ and $\delta q \sim C_q e^{3u} $, or in terms of $\rho$,
	\begin{align}
	\delta p &\sim C_p \rho^{5/2} & \delta q &\sim C_q \rho^{-3/2} 
	\end{align}
These asymptotic forms are consistent with the scaling dimensions in (\ref{scalingdims2}), as we either have solutions that scale as $\rho^\Delta$ or $\rho^{2 - \Delta}$.
We see that the $q$ scalar diverges as we approach the $\cN=(1,1)$ point as $\rho \to 0$ unless we fine-tune the coefficient $C_q$ to zero.
This is identified with turning off the source for an operator with  scaling dimension  larger than 2 on the boundary CFT.

A similar counting as before shows that the flow preserves two of the four supersymmetries of the $\cN=(1,1)$ vacuum.
Therefore, we have RG-flow interfaces between a CFT with central charge $c^{(4,4)}$ and a CFT with central charge  $c^{(1, 1)}$, where \cite{Brown:1986nw, Henningson:1998gx}
	\begin{align}
	\frac{c^{(1,1)}}{c^{(4,4)}} = \sqrt{ \frac{W^{(4,4)}_0}{W^{(1,1)}_0} } = \frac{1}{2}
	\end{align}

\subsection{Truncation 3}

\begin{figure}[htbp]
\begin{center}
{\centerline{\includegraphics[width=10cm]{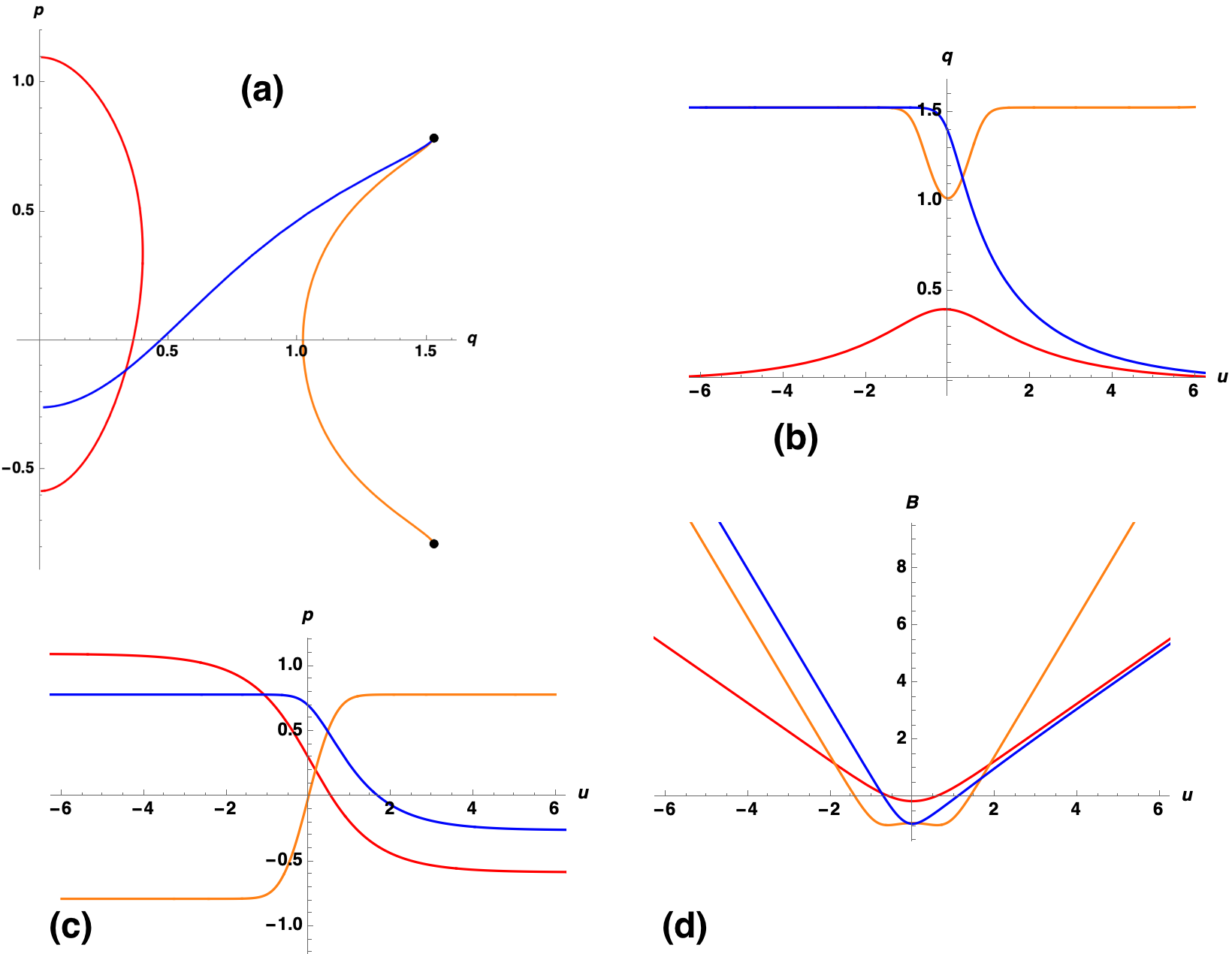}}}
\caption{(a) $pq$  parametric plot, (b) plot of $q$, (c) plot of $p$, (d) plot of the metric function $B$ as functions of the $\AdS_2$ slicing coordinate $u$ for truncation 3.}
\label{fig:fig3}
\end{center}
\end{figure}

The analysis of the flow equations and their solutions for truncation 3 proceeds very similarly to the one for truncation 2, presented in the previous section. 
The matrix $A_1$ takes the form,
	\begin{align}
	A_1 = \left( \resizebox{0.8\textwidth}{!}{ $\begin{array}{cccccccc}
	0 & 0 & 0 & (a_1 - a_2 + b_2) \cos p & 0 & (-a_1 + a_2 + b_1) \sin p & 0 & 0 \\ 
	0 & -2 a_1 \cos p & (a_1 + a_2 - b_2) \cos p & 0 & (a_1 + a_2 - b_1) \sin p & 0 & 0 & 2 a_2 \sin p \\
	0 & (a_1 + a_2 - b_2) \cos p & -2 a_1 \cos p & 0 & -2 a_2 \sin p & 0 & 0 & (-a_1 - a_2 + b_1) \sin p \\
	(a_1 - a_2 + b_2) \cos p & 0 & 0 & 0 & 0 & 0 & (a_1 - a_2 - b_1) \sin p & 0 \\
	0 & (a_1 + a_2 - b_1) \sin p & -2 a_2 \sin p & 0 & 2 a_1 \cos p & 0 & 0 & (a_1 + a_2 - b_2) \cos p \\
	(-a_1 + a_2 + b_1) \sin p & 0 & 0 & 0 & 0 & 0 & (a_1 - a_2 + b_2) \cos p & 0 \\
	0 & 0 & 0 & (a1 - a2 - b1) \sin p & 0 & (a_1 - a_2 + b_2) \cos p & 0 & 0 \\
	0 & 2 a_2 \sin p & (-a_1 - a_2 + b_1) \sin p  & 0 & (a_1 + a_2 - b_2) \cos p & 0 & 0 & 2 a_1 \cos p
	\end{array}$ } \right)
	\end{align}
where we define, for this truncation,
	\begin{align}
	a_1 &= \frac{1}{2} (1 + \cosh q) \cosh q \sin^2 p~, & a_2 &= \frac{1}{2} (1 + \cosh q) \cosh q \cos^2 p \nonumber \\
	b_1 &= \frac{1}{2} (1 + \cosh q) (1 + \cosh^2 q) \sin^2 p ~, & b_2 &= \frac{1}{2} (1 + \cosh q) (1 + \cosh^2 q) \cos^2 p 
	\end{align}
As with trunctation 2, there are two eigenvalues of $(A_1)^2$: one with six-fold degeneracy that does not satisfy (\ref{susyvac}), and one with two-fold degeneracy that does. 
This eigenvalue is
	\begin{align}
	w^2 = \frac{1}{64} \cosh^4 \frac{q}{2} \qty(175 - 224 \cosh q + 140 \cosh 2q - 32 \cosh 3q + 5 \cosh 4q + 24 \cos 4p \sinh^4 q )
	\end{align}
The corresponding eigenvectors of $A_1$ with eigenvalue $\pm w$ are
	\begin{align}
	v_\pm &= \qty\big(-(3 a_1 + a_2 - b_2) \cos p \pm w) v_1 + (a_1 + 3 a_2 - b_1) \sin p ~ v_2
	\end{align}
where $v_1, v_2$ are defined as before in (\ref{v1v2}).
The flow equations once again do not take a simple form and must be solved numerically.
In figure \ref{fig:fig3} we present some examples for the numerical solutions of the flow equations for truncation 3, which exhibit very similar features to the solutions of the flow equations for truncation 2. 
By fine-tuning initial conditions we can produce    flows that (i) look like the Janus solutions in truncation 1, plotted in red in  figure \ref{fig:fig3}, (ii) connect $\cN=(4,4)$ and $\cN=(1,1)$ vacua,  plotted in blue in  figure \ref{fig:fig3}, and (iii) connect two $\cN=(1,1)$ vacua which are related by flipping signs of $p$, plotted in orange in  figure \ref{fig:fig3}. In the $pq$ parametric  plot given  in figure \ref{fig:fig3}(a), the location of the $\cN=(1,1)$ vacua  $p=\pm {\pi\over 4}$, $q= \sinh^{-1} \sqrt{2+2\sqrt{2}}$ are denoted by black dots.

The $\cN = (1, 1)$ points are again repulsive.
With $g=1/2$, the $\cN = (1, 1)$ vacuum potential is $W_0 = (1 + \sqrt{2})^2/2$ and the $\AdS_3$ length scale is $L = \sqrt{2} - 1$.
The linear expansion around the vacuum yields the following mass-squares for $\delta p$ and $\delta q$ fluctuations,
	\begin{align}
	m_p^2L^2 &= 1~, \qquad m_q^2L^2 = 2 + 2\sqrt{2} 
	\end{align}
which correspond to the scaling dimensions,
	\begin{align} \label{scalingdims3}
	\Delta_p &= 1 + \sqrt{2} ~, \qquad \Delta_q = 2+\sqrt{2}
	\end{align}
The linearized flow equations around the $\cN=(1,1)$ point are
	\begin{align}
	\delta p' &= -(3 + 2\sqrt{2}) \delta p + \cdots \nonumber \\
	\delta q' &= (2+\sqrt{2}) \delta q + \cdots
	\end{align}
which are solved by $\delta p \sim C_p e^{-(3 + 2\sqrt{2})u}$ and $\delta q \sim C_q e^{(2+\sqrt{2})u} $, or in terms of $\rho$ by substituting $e^u \sim \rho^{-L}$,
	\begin{align}
	\delta p &\sim C_p \rho^{1 + \sqrt{2}} & \delta q &\sim C_q \rho^{-\sqrt{2}} 
	\end{align}
These asymptotic forms are consistent with the scaling dimensions in (\ref{scalingdims3}).
Again, we see that the $q$ scalar diverges as $\rho \to 0$ unless we fine-tune the coefficient $C_q$ to zero, which turns off the source for an operator with  scaling dimension  larger than 2 on the boundary CFT.

A similar counting as before shows that the flow preserves two of the four supersymmetries of the $\cN=(1,1)$ vacuum.
Therefore, we have RG-flow interfaces between a CFT with central charge $c^{(4,4)}$ and a CFT with central charge $c^{(1,1)}$, where
	\begin{align}
	\frac{c^{(1,1)}}{c^{(4,4)}} = \sqrt{ \frac{W^{(4,4)}_0}{W^{(1,1)}_0} } = \sqrt{2} - 1
	\end{align}

\section{Discussion}\label{sec5}
In the paper, we constructed new solutions of three-dimensional gauged supergravity.  The solutions produced describe interface CFTs holographically.  We considered three different truncations of the scalar fields which are associated with three different supersymmetric $\AdS_3$ vacua with $\cN=(4,4)$ and $\cN=(1,1)$ supersymmetry. The solution in the first truncation is a Janus solution that is very similar to the one found in \cite{Chen:2020efh} for a simpler three-dimensional gauged supergravity.  The CFTs on both sides of the interface are deformations of the $\cN=(4,4)$ vacuum with a source for a marginal operator turned on as well as a position-dependent expectation value for a relevant operator. The interface preserves half of the sixteen supersymmetries of the $\cN=(4,4)$ vacuum.
	
	The solutions for the other two truncations represent RG-flow interfaces in the sense that the solutions we find have  different CFTs on each side of the interface. For example, we find solutions where the interface connects the $\cN=(4,4)$ CFT  (with a marginal operator and relevant expectation value) and the $\cN=(1,1)$ CFT (with an irrelevant source). Note that  there is no clear distinction between the UV and the IR in the RG-flow Janus solutions,  since both sides put together form the boundary of the asymptotically AdS space. This is to be contrasted with a Poincar\'e-sliced RG-flow solution, where the AdS boundary with the larger  curvature radius (or central charge) is viewed as describing the UV CFT.  A irrelevant source is turned on near the $\cN=(1,1)$ asymptotic AdS, which means that, from the perspective of the flow, this constitutes a repulsive direction. To find a flow that comes very close to the $\cN=(1,1)$ vacuum, we have to fine-tune our initial conditions, which corresponds to fine-tuning the source of the irrelevant operator. 
	
	We have worked in truncations where the dynamics of the eight scalar fields $q_i, p_i$ for $i=1,2,3,4$ are reduced to the dynamics of two scalars $q,p$, where in all three cases $q=0$ corresponds to the $\cN=(4,4)$ vacuum. Hence, we could find interface CFTs of the $\cN=(4,4)$ CFT with one of the $\cN=(1,1)$ CFTs. In this truncation, we cannot find an interface solution connecting  the two distinct $\cN=(1,1)$ vacua. For such a solution, we would have to consider the flow equations with at least four independent scalars. The fine-tuning of the initial conditions to produce the interface solution would  also be more challenging.  
	
	The $\SO(4) \times \SO(4)$ gauging depends on a real parameter $\alpha$ and in this paper we have only considered the case $\alpha=1$ which simplifies  the expression of the $A_i$ matrices and the scalar potential. We expect that the solutions for other choices of $\alpha$  behave qualitatively the same, since the supersymmetric vacua exist for other values of $\alpha$.
	Nevertheless, it would be interesting to check this  expectation.  It would also be interesting to consider holographic observables such as the entanglement entropy around the interface or correlation functions. We plan to investigate   some of these questions in the future.

\section*{Acknowledgements}

The work of M.~G.~was supported, in part, by the National Science Foundation under grant PHY-19-14412. 
The authors are grateful to the Mani L.~Bhaumik Institute for Theoretical Physics for support.

\newpage
\appendix

\section{$\SO(8)$ Gamma matrices}\label{appendix-gamma}

We are working with $8 \times 8$ Gamma matrices $\Gamma^I_{A \dot A}$ and their transposes $\Gamma^I_{\dot A A}$, which satisfy the Clifford algebra,
	\begin{align} \Gamma^I_{A \dot A} \Gamma^J_{\dot A B} + \Gamma^J_{A \dot A} \Gamma^I_{\dot A B} = 2 \delta^{IJ} \delta_{AB}
	\end{align}
Explicitly, we use the basis in Green-Schwarz-Witten \cite{Green:1987sp},
	\begin{align}
	\Gamma^8_{A \dot A} &= 1 \otimes 1 \otimes 1~, &  \Gamma^1_{A \dot A} &= i \sigma_2 \otimes i \sigma_2 \otimes i \sigma_2  \nonumber \\
	\Gamma^2_{A \dot A} &= 1 \otimes \sigma_1 \otimes i\sigma_2~, &  \Gamma^3_{A \dot A} &= 1 \otimes \sigma_3 \otimes i\sigma_2 \nonumber \\
	\Gamma^4_{A \dot A} &= \sigma_1 \otimes i\sigma_2 \otimes 1~, &  \Gamma^5_{A \dot A} &= \sigma_3 \otimes i\sigma_2 \otimes 1  \nonumber \\
	\Gamma^6_{A \dot A} &= i\sigma_2 \otimes 1 \otimes \sigma_1~, &  \Gamma^7_{A \dot A} &= i\sigma_2 \otimes 1 \otimes \sigma_3
	\end{align}
The matrices $\Gamma^{IJ}_{AB}$, $\Gamma^{IJ}_{\dot A \dot B}$ and similar are defined as antisymmetrized products of $\Gamma$s with the appropriate indices contracted.
For instance,
	\begin{align} \Gamma^{IJ}_{AB} \equiv \frac{1}{2} (\Gamma^I_{A \dot A} \Gamma^J_{\dot A B} - \Gamma^J_{A \dot A} \Gamma^I_{\dot A B}) 
	\end{align}

\newpage

\providecommand{\href}[2]{#2}\begingroup\raggedright\endgroup

\end{document}